\documentclass[12pt]{article}
\usepackage{epsfig}

\newcommand{\be}{\begin{equation}}
\newcommand{\ee}{\end{equation}}
\newcommand{\ba}{\begin{eqnarray}}
\newcommand{\ea}{\end{eqnarray}}

\newcommand{\al}{\alpha}
\newcommand{\alp}{\frac{\alpha}{\pi}}

\newcommand{\ice}[1]{\relax}
\begin{document}
\thispagestyle{empty}
\begin{flushright}
MZ-TH/01-28\\
October 2001\\
\end{flushright}
\vspace{0.5cm}
\begin{center}
{\Large\bf Muon anomalous magnetic moment:\\ a consistency check
for the next-to-leading order hadronic contributions }\\[1truecm]
{\large \bf A.A.Pivovarov}

Institut f\"ur Physik, Johannes-Gutenberg-Universit\"at,\\
Staudinger Weg 7, D-55099 Mainz, Germany\\
and \\
Institute for Nuclear Research of the\\
Russian Academy of Sciences, Moscow 117312, Russia
\end{center}

\vspace{1truecm}
\begin{abstract}
\noindent A model for verifying a consistency of the 
next-to-leading order hadronic contributions to the muon anomalous 
magnetic moment with those of the leading order is proposed.
A part of the next-to-leading order hadronic contributions related to 
the vacuum polarization is rather accurately reproduced in the model.
I find a new numerical value for the light-by-light hadronic contribution
that leads to agreement with recent experimental result 
for the muon anomalous magnetic moment.
\end{abstract}

\newpage
\section{Introduction}
A numerical value of the muon anomalous magnetic moment (MAMM) measured
experimentally with high precision can be used to quantitatively test 
the theories suggested for describing particle interactions. 
The experimental result for MAMM presented in ref.~\cite{czamar} reads 
\be
\label{exp}
a_\mu^{\rm exp}=116~592~023(151)\times 10^{-11}
\ee
with the uncertainty $151\times 10^{-11}$. 
The main anomalous effect is due to Schwinger term
\be
\label{Schw}
a_\mu^{\rm Schw}=\frac{\al}{2\pi}
\ee
where $\al$ is the fine structure constant $\al^{-1}=137.036\dots$
The theoretical contributions presently 
computed in the standard model for the comparison with 
the experimental value given in eq.~(\ref{exp}) are divided
into three parts: leptonic (QED), electroweak (EW), and hadronic (had)
one. The pure leptonic part is computed in perturbative QED 
through $\al^5$ order~\cite{tenthQED,qedcont}.
The numerical value of the QED contribution
to the muon anomalous magnetic moment reads (as a review see~\cite{czamar})
\be
\label{QED}
a_\mu^{\rm QED}=116~584~705.7(2.9)\times 10^{-11}\, .
\ee
The EW corrections are well defined in the perturbation theory framework 
of the standard model and have been computed with the two-loop accuracy
(as a review see~\cite{czamar})
\be
\label{EW}
a_\mu^{\rm EW}=152(4)\times 10^{-11}\, . 
\ee
Numerically this contribution matches the present 
experimental uncertainty. The EW correction 
will be noticeable if a goal to reach the 
planned experimental accuracy $40\times 10^{-11}$
is accomplished (as a review see~\cite{czamar}).

The hadronic contribution to MAMM is sensitive to the infrared region 
and cannot be computed in perturbative QCD with light quarks. 
The current masses of
light quarks are too small to provide a necessary infrared cutoff and 
explicit models of hadronization are required for the quantitative
analysis. This constitutes 
a main difficulty of the theoretical analysis of MAMM in the
standard model.
Writing
\be
\label{SM}
a_\mu^{\rm SM}=a_\mu^{\rm QED}+a_\mu^{\rm EW}+a_\mu^{\rm had}
\ee
and assuming 
\be
\label{assumption}
a_\mu^{\rm exp}=a_\mu^{\rm SM}
\ee
one finds a numerical value for the hadronic contribution
to MAMM
\be
\label{th}
a_\mu^{\rm had}|_{\rm th}
=(7165.3+151|_{\rm exp} + 2.9|_{\rm QED}+4|_{\rm EW})\times 10^{-11}
\ee
with the experimental error dominating the uncertainty.

Since the hadronic contribution is sensitive to the details of the 
strong coupling regime of QCD at low energies and cannot be unambiguously 
computed in perturbation theory framework the theoretical prediction 
for MAMM in the standard model depends crucially on how this
contribution is estimated. In the absence of reliable theoretical 
tool for computation in this region one turns to experimental data  
on low-energy hadron interactions for extracting a necessary numerical value.
In general terms the hadronic contribution to MAMM is determined
by the correlation functions of electromagnetic (EM) currents. 
As the source for the EM current is readily available for a wide range 
of energies one tries to extract these functions
(or some their characteristics) from experiment.
Without an explicit use of QCD the correction $a_\mu^{\rm had}$
is generated through the EM interaction $e j_\mu^{had}A^\mu$ 
with $j_\mu^{had}$ being the hadronic part of the EM current.
At the leading order ($\alpha^2$ in the formal power-counting)
only the two-point correlation function of the EM currents emerges
in the analysis of hadronic contributions to MAMM
\be
\label{pi2}
\Pi_2\sim\langle j_\mu^{had}(x)j_\nu^{had}(0)\rangle \, .
\ee
At the next-to-leading order ($\alpha^3$)
the four-point correlation function appears
\be
\label{fourcorr}
\Pi_4\sim
\langle j_\mu^{had}(x)j_\nu^{had}(y)
j_\al^{had}(z)j_\beta^{had}(0)\rangle \, .
\ee
These correlators are not calculable perturbatively in the region
that is essential for the determination of the hadronic contributions 
to MAMM. The leading contribution to MAMM comes 
from the two-point correlator eq.~(\ref{pi2}) referred 
to as the hadronic vacuum polarization contribution
while the four-point function eq.~(\ref{fourcorr})
first emerges at the $\al^3$ order, 
most explicitly as the light-by-light scattering.
To avoid using QCD in the strong coupling mode
one has to extract the necessary contribution to MAMM
by studying these two correlation
functions experimentally without an explicit realization 
of the hadronic EM current $j_\mu^{had}$ in terms of elementary fields.
Historically this was a way of studying the EM properties of hadrons
before emerging QCD as a fundamental theory of
strong interactions~(e.g.~\cite{sakurai}).

\section{Hadronic contribution at the leading order}
At the leading order in $\al$ the hadronic contribution 
is described by the correlator 
\be
\label{vacpolin}
i\int \langle Tj_\mu^{had}(x)j_\nu^{had}(0) \rangle e^{iqx}dx=
(q_\mu q_\nu -g_{\mu\nu}q^2)\Pi^{\rm had}(q^2)
\ee
which reduces to a single function $\Pi^{\rm had}(q^2)$ 
of one variable $q^2$. The correlator is transverse 
due to conservation of the hadronic EM current in the standard model. 
This function gives a contribution
to the muon anomalous magnetic moment~(e.g. \cite{leadorder})
\be
\label{directcont}
a_\mu^{\rm had}({\rm LO})
=4\pi\left(\frac{\al}{\pi}\right)^2\int_{4m_\pi^2}^\infty
\frac{ds}{s}K(s){\rm Im}~\Pi^{\rm had}(s)
\ee
with the one-loop kernel of the form
\be
\label{oneloopkern}
K(s)=\int_0^1dx\frac{x^2(1-x)}{x^2+(1-x)\frac{s}{m_\mu^2}}\, .
\ee
Here 
${\rm Im}~\Pi^{\rm had}(s)={\rm Im}~\{\Pi^{\rm had}(q^2)|_{q^2=s+i0}\}$,
$m_\mu$ is a muon mass. 

The leading order hadronic contribution to MAMM
is represented by an integral over the hadron spectrum. 
No specific information about the function 
${\rm Im}~\Pi^{\rm had}(s)$ is necessary point-wise except 
its threshold structure in the low-energy region.   
For the applications at the leading order in $\al$ the function 
${\rm Im}~\Pi^{\rm had}(s)$
can be uniquely identified with data from $e^+e^-$ annihilation
into hadrons. Introducing the experimental $R^{\rm exp}(s)$ ratio
\be
\label{expR}
R^{\rm exp}(s)
=\frac{\sigma(e^+e^-\to {\rm hadrons})}
{\sigma(e^+e^-\to \mu^+\mu^-)}\, ,
\qquad s=(p_{e^+}+p_{e^-})^2
\ee
and identifying it with the theoretical quantity
$R^{\rm th}|_{LO}(s)$ taken at the leading order in $\al$ 
as
\be
R^{\rm th}|_{LO}(s)=12\pi {\rm Im}~\Pi^{\rm had}(s)
\ee
one finds 
\be
\label{exp0vac}
a_\mu^{\rm had}({\rm LO})
=\frac{1}{3}\left(\frac{\al}{\pi}\right)^2\int_{4m_\pi^2}^\infty
\frac{R^{\rm exp}(s)K(s)}{s}ds\, .
\ee
The contribution to MAMM based on the representation given 
in eq.~(\ref{exp0vac}) is well studied. Several recent 
determinations are 
\be
\label{kinosh}
a_\mu^{\rm had}({\rm LO})=7011(94)\times 10^{-11}
\qquad{\rm (ref.~\cite{kinohad})};
\ee
\be
\label{vacpolDH}
a_\mu^{\rm had}({\rm LO})=6924(62)\times 10^{-11}
\qquad{\rm (ref.~\cite{davhock})};
\ee
\be
\label{jeger}
a_\mu^{\rm had}({\rm LO})=6988(111)\times 10^{-11}
\qquad{\rm (ref.~\cite{jeger})}\, .
\ee
I use in the further analysis a naive average of these three results 
(both central values and errors are averaged) which reads
\be
\label{piv}
a_\mu^{\rm had}({\rm LO})=6974(89)\times 10^{-11}\, .
\ee
Writing 
\[
a_\mu^{\rm had}|_{\rm th}= a_\mu^{\rm had}({\rm LO})
+ a_\mu^{\rm had}({\rm NLO})
\]
and comparing with eq.~(\ref{th}) one has (in units $10^{-11}$)
\[
a_\mu^{\rm had}({\rm NLO}) =
7165+151|_{\rm exp} + 2.9|_{\rm QED}+4|_{\rm EW}-6974(89)|_{\rm LO}
\]
\be
=191+151|_{\rm exp} + 2.9|_{\rm QED}+4|_{\rm EW}+89|_{\rm had}\, .
\ee
Assuming the statistical independence of the uncertainties 
one finds after adding them in quadratures
\be
\label{nlopres}
a_\mu^{\rm had}({\rm NLO}) = (191\pm 175)\times 10^{-11}
\ee
that does not allow to see higher order hadronic 
effects clearly. The error comes mainly from the experimental value  
eq.~(\ref{exp}) and the leading order hadronic data eq.~(\ref{piv})
the statistical correlation of which is supposed to be small
as they come from different sources. Other errors are negligible.
For the target experimental error of MAMM at the level of
$40\times 10^{-11}$ one finds the uncertainty 
of the NLO hadronic contribution to become 
$98\times 10^{-11}$. Assuming that the mean value of 
$a_\mu^{\rm exp}$ in the planned experiment will not change 
one finds a numerical value for the NLO hadronic contribution 
\be
\label{futureNLOexp} 
a_\mu^{\rm had}({\rm NLO}) = (191\pm 98)\times 10^{-11}
\ee
that makes the NLO hadronic effects noticeable at the level 
of two standard deviations. 
If the mean value of $a_\mu^{\rm exp}$
will change in the range of the present experimental
uncertainty $151\times 10^{-11}$ 
the NLO hadronic effects can be more or less pronounced.
From the naive counting in $\al$ a numerical value 
for the theoretical NLO hadronic contribution about $50\times 10^{-11}$
can be expected. This number is comparable in magnitude
with the uncertainty in eq.~(\ref{futureNLOexp}) 
and should be taken into account.

\section{Hadronic contribution at next-to-leading order}
In NLO there is no such a transparency with determining hadronic
contributions as in LO. 
Basically there are two new features. 
On the experiment side the interpretation of data to be used
in the NLO theoretical calculations is more involved. The problem is
to avoid double counting as a part of the hadronic 
contributions has already been accounted through the use of data at LO.
On the theory side a new correlation function $\Pi_4$ from 
eq.~(\ref{fourcorr}) which is much more
complicated than the two-point correlator enters the game.
At present there is no accurate experimental 
determination of the four-point function in the kinematical range 
necessary for MAMM computation and one has to rely on phenomenological
models used for this function. It is difficult to control 
the accuracy of such models that introduces an explicit model dependence in
the calculation of the NLO hadronic contribution and makes predictions
less definite than in LO.

\subsection{Interpretation of data at NLO of EM interaction}
For applications at NLO in $\al$ the extraction of data is more
involved. For instance, one should explicitly take into account the NLO
corrections to theoretical factors that emerge in a description of 
the process from which a set of data is taken. 
These ``theoretically corrected'' data should be used in NLO
applications for computing MAMM. As the sets of data are mainly
extracted from $e^+e^-$ annihilation I discuss this particular process
in some detail.

\subsubsection{One-photon mediated $e^+e^-$ annihilation} 
The main object studied experimentally in this sector
is the full photon propagator $D(q^2)$
\be
D(q^2)=\frac{1}{-q^2}\frac{1}{1+e^2\Pi(q^2)}
\ee
with $\Pi(q^2)=\Pi^{lept}(q^2)+\Pi^{had}(q^2)$
being a one-particle irreducible block, $e^2=4\pi \al$. 
Note that in higher orders the one-particle irreducible block
does not split into a sum of pure leptonic and pure hadronic contributions.
It happens first at NNLO which is far beyond the practical interest
though. I discuss only NLO or $\al^3$ terms in the formal $\al$
power-counting. Since the data are collected at low energies 
the EW sector can be excluded.
With these restrictions the cross section 
of $e^+e^-$ annihilation through the one-photon exchange at NLO
without vertex corrections to initial states is proportional to 
\[
{\rm Im}~\{q^2 D(q^2)|_{q^2=s+i0}\} = \frac{e^2 {\rm Im}\Pi(s)}
{(1+e^2 {\rm Re} \Pi(s))^2+e^4 {\rm Im} \Pi(s)^2}
\]
\be
\label{den}
=\frac{e^2({\rm Im} \Pi^{\rm lept}(s)+{\rm Im} \Pi^{\rm had}(s))}
{(1+e^2 {\rm Re} \Pi(s))^2+e^4 {\rm Im} \Pi(s)^2}\, .
\ee
The theoretical expression for the $R$ ratio at NLO reads
\be
\label{rNLO}
R^{\rm th}|_{NLO}(s)
=\frac{{\rm Im} \Pi^{\rm had}(s)}{{\rm Im} \Pi^{\mu\mu}(s)}\, .
\ee
If $R^{\rm th}|_{NLO}(s)$ is identified with $R^{\rm exp}(s)$
from eq.~(\ref{expR})
then ${\rm Im} \Pi^{\rm had}(s)$
can be restored by using a theoretically calculated 
${\rm Im} \Pi^{\mu\mu}(s)$.
For $s\gg m_\mu^2$ one finds with NLO accuracy
\be
\label{expRmm}
12\pi {\rm Im}\Pi^{\rm had}(s)
=R^{\rm exp}(s)~\left(1+\frac{3}{4}\frac{\al}{\pi}\right)\, .
\ee
In some analyses the cross section 
$\sigma(e^+e^-\to {\rm hadrons})$ 
divided by the normalization factor 
\be
\sigma_0=\frac{4\pi\al^2}{3s}
\ee
is used as a data set~\cite{czamar}.
Then the relation 
\be
12\pi {\rm Im} \Pi^{\rm had}(s)
=\sigma(e^+e^-\to {\rm hadrons})/\sigma_0+O(\al)
\ee
is valid only at the leading order in $\al$.
One of the differences with the $R$ ratio at NLO from eq.~(\ref{rNLO})
is the term ${\rm Re} \Pi^{\rm had}(s)$ from 
the denominator in eq.~(\ref{den}).
The quantity ${\rm Re} \Pi^{\rm had}(s)$
can be found by reiterating the leading order term 
${\rm Im} \Pi^{\rm had}(s)$ through the 
dispersion relation that gives a relative
error of $\al^2$ order. The NLO contribution in the denominator
is related to the running of the EM coupling constant and can partly
be taken into account through the renormalization group technique
for the energies far from the resonances~\cite{eucl}. Another
difference is the corrections
to the production vertex that should properly be taken into account 
as they enter the cross section. Extracting  
${\rm Im} \Pi^{\rm had}(s)$ from the cross section requires to 
subtract these corrections from the data in order 
to avoid double counting in the NLO analysis of MAMM 
if a theoretical NLO kernel for averaging 
the two-point correlator is used.

The use of the $R$-ratio is 
preferable from the theoretical point of view as it relates data to 
the imaginary part of the two-point hadronic correlator  
${\rm Im} \Pi^{\rm had}(s)$ in a simple way.
It is also preferable from the experimental point of view 
since the total normalization of the data is fixed that helps to 
eliminate systematic errors. 
In this respect a $\tau$-data set that can be used to determine
that part of the two-point function that is generated by the isovector
part of the hadronic EM current
in the limit of exact isotopic invariance 
has a different normalization at
NLO and should be corrected by an explicit account for 
contributions of the relative $\al$ order.
Note that the NLO correction emerging from the interpretation of data
can be controlled theoretically within counting in $\al$
while corrections due to the violation of isotopic invariance
for the data obtained from the $\tau$ and $e^+e^-$ channels
can only be estimated in models.
The problem of different normalization remains also for heavy hadrons
if their contribution to the cross section is calculated from their
leptonic branchings. The NLO contribution of heavy flavors 
is not essential though because it is small.

\subsubsection{Two-photon mediated $e^+e^-$ annihilation}
The NLO cross section of $e^+e^-$ annihilation
contains a contribution of the two-photon 
annihilation with one hadronic insertion into the photon propagator.
This contribution requires a special treatment 
before the data set is related to
the hadronic two-point function with the NLO accuracy.
For instance, the NLO kernel for the MAMM diagram with a vertex correction
integrates the part of data emerging through the double photon scattering
channel in $e^+e^-$ annihilation.
This can lead to double counting at NLO for MAMM.

Thus, one concludes that at NLO a strict correlation between sets
of data and theoretical expressions for the NLO kernels emerges.
This leads to additional contributions of the relative $\al$ order 
that numerically amounts to about 1\% of the leading order
contribution which is the precision one is trying to reach for the 
comparison with the experimental result for MAMM.

\subsection{Four-point correlator}
At NLO a new correlation function of hadronic EM currents emerges.
This correlation function leads to a new effect which is known
as light-by-light scattering.
Besides this explicit effect a less pronounced mixed effect also emerges.
The four-point function gives a contribution to the full photon 
propagator of the form 
\be
\label{propprojph0}
\int dxdy D_{\mu\nu}(x-y)
\langle Tj_\mu^{had}(x)j_\nu^{had}(y)j_\al^{had}(z)j_\beta^{had}(0)\rangle
\ee
where $D_{\mu\nu}(x)$ is a free photon propagator
with a scalar amplitude $D(x)\sim 1/x^2$.
In other words a projection of the four-point function of the 
form 
\be
\label{proj2phot}
\int \frac{dxdy}{(x-y)^2}
\langle Tj_\mu^{had}(x)j_\mu^{had}(y)j_\al^{had}(z)j_\beta^{had}(0)\rangle
\ee
is present in the two-photon Green function.
In QCD and other models where the EM current 
is explicitly expressed through the
elementary fields this contribution is interpreted as an EM
correction to the one-particle irreducible block.
There is an option to include this contribution to the two-point
function. I do not consider this option since the picture of the local
interaction of the photon with the hadronic EM current is lost in such case.

Thus, an accurate account of NLO hadronic contributions to MAMM
from general principles is rather a challenging task both
experimentally and theoretically. As a first approach to it 
one can use an effective theory with few free parameters 
providing a unique framework for calculations 
at LO and NLO. In such an approach the LO information is used 
to obtain numerical values for the model parameters.
The NLO results are then computed theoretically. This approach 
can also serve as a base for verifying a consistence of the estimates
for the NLO hadronic contributions 
made in different phenomenological models.

\section{A model for hadronic contributions}
In this section I describe a model to check a consistence of 
the NLO hadronic contributions and especially the light-by-light 
contribution with the results of LO analysis for MAMM.
The simplest version of the model contains 
three light quarks with QCD quantum numbers
and the mass $m_q$ which is the only 
model parameter. The numerical value of $m_q$ 
is fixed from the LO hadronic contribution and then
used to find the NLO result. Heavy quarks enter the model 
with their standard masses.
In this model the calculations are explicit and can be performed
analytically that is an advantage. Indeed, the model 
differs from the leptonic sector only by the QCD group factors and
the numerical values of fermion masses.

\subsection{Fixing $m_q$ from the LO hadronic contribution}
A fermion with mass $m_q$ without QCD group factors 
(as a lepton) gives the LO contribution to MAMM 
of the form
\be
\label{mod0vac}
a_\mu^{\rm ferm}({\rm LO})
= I(m_q)\left(\frac{\al}{\pi}\right)^2
\ee
with 
\be
I(m_q)=\int_{4m_q^2}^\infty \frac{\rho_q(s)K(s)}{s}ds
\ee
and
\be
\label{fermspect}
\rho_q(s)=\frac{1}{3}
\sqrt{1-\frac{4 m_q^2}{s}}\left(1+\frac{2 m_q^2}{s}\right)\, .
\ee
Explicit integration over $s$ with the kernel $K(s)$ from 
eq.~(\ref{oneloopkern}) gives
\be
\label{intrepI}
I(m_q)=\int_0^1 dx (1-x) [-\pi(x,m_q)]
\ee
where 
\be
\pi(x,m_q)=\left(\frac{1}{3z}-1\right)\varphi(z)-\frac{1}{9}
\ee
and
\be
\label{lasttech}
\varphi(z)=\frac{1}{\sqrt{z}}{\rm ArcTanh}(\sqrt{z})-1,\quad
z=\frac{m_\mu^2 x^2}{4m_q^2(1-x)+ m_\mu^2 x^2}\, .
\ee
An analytical expression for the function $I(m_q)$ is known, however,
the integral representation given in eq.~(\ref{intrepI})
is sufficient for practical applications.

Contributions of the heavy $c$ and $b$ quarks
can directly be computed in QCD perturbation theory
independently of the model. In the present calculation I use only 
free quark approximation for simplicity.
For the $c$ quark with the mass $m_c=1.6~{\rm GeV}$~\cite{PDG} 
and the charge $e_c=2/3$ one finds from eq.~(\ref{mod0vac}) multiplied
by the group factor $3 e_c^2=4/3$ 
\be
a_\mu^{\rm mod}({\rm LO; c})= 69.3\times 10^{-11} \, .
\ee
The $b$-quark contribution for $m_b=4.8~{\rm GeV}$~\cite{mb} 
and $e_b=-1/3$ is small and reads
\be
a_\mu^{\rm mod}({\rm LO; b})= 1.9\times 10^{-11} \, .
\ee
Thus the contribution of light hadronic modes that is represented in our model 
by light fermions with the mass $m_q$ amounts to 
\be
\label{vacpoluds}
a_\mu^{\rm mod}({\rm LO; uds})
=(6974.3-69.3-1.9)\times 10^{-11}
=6903(89)\times 10^{-11}\, .
\ee
I assume that this result directly
corresponds to the contribution of the two-point correlator at 
the leading order as given in eq.~(\ref{directcont}).
It means that a real data set is properly corrected to extract 
${rm Im} \Pi^{\rm had}(s)$. As was discussed above the
extraction of ${rm Im} \Pi^{\rm had}(s)$ with the NLO accuracy
requires
a careful interpretation of data which is assumed to be done.

From eqs.~(\ref{mod0vac}-\ref{lasttech}) and eq.~(\ref{vacpoluds}) 
one obtains a numerical value for the single model parameter $m_q$ 
\be
\label{mqval}
m_q=179\pm 1~{\rm MeV}\, .
\ee 
This value is rather close to that of the charged 
pion mass, $m_\pi=139.6~{\rm MeV}$, which is
expected as the LO contribution is mainly sensitive to the first
derivative at $q^2=0$
of the two-point function $\Pi^{\rm had}(q^2)$ given in
eq.~(\ref{vacpolin}). For definiteness I give the LO contribution of
the light hadronic modes within the model
obtained literally with the number from eq.~(\ref{mqval})
\be
\label{vacpoludsback}
a_\mu^{\rm mod}({\rm LO; uds})(m_q=179\pm 1~{\rm MeV})
=(6920\pm70)\times 10^{-11}\, .
\ee
Since in the framework of the model the NLO hadronic corrections to MAMM
are determined by the single parameter $m_q$ with the numerical value
from eq.~(\ref{mqval}) they can readily be found.

\subsection{Hadronic contributions at NLO}
The first check is to use the model for computing 
the higher order hadronic corrections due to vacuum polarization graphs.
The data-based analysis gives for 
the NLO effects of this type~\cite{krause}
\be
\label{vacpol2}
a_\mu^{\rm had}({\rm vac;NLO})=-101(6)\times 10^{-11}\, .
\ee
This number is about 1.5\% of the leading term as expected.
As was discussed above at this level of precision
the numerical value for the NLO contribution
depends strongly on the data sets used in the analysis.
For different data sets the different expressions of the NLO kernel should be
used to avoid double counting. For example, if the $R$-ratio is used
in the one-loop computation then the LO result should first 
be divided by the factor (cf. eq.~(\ref{expRmm}))
\be
12\pi {\rm Im}\Pi^{\mu\mu}(s)|_{s\gg m_\mu^2}
=\left(1+\frac{3}{4}\frac{\al}{\pi}
+O\left(\frac{m_\mu^4}{s^2}\right)\right)
\ee
before being used in the NLO analysis that changes 
the LO result by $12\times 10^{-11}$ that exceeds the uncertainty quoted 
in eq.~(\ref{vacpol2}). In fact, even mass suppressed terms can be
important at this level of precision
and the entire function ${\rm Im}\Pi^{\mu\mu}(s)$ 
should be integrated since the mass terms from the leading order 
can partly cancel the NLO corrections in $\al$.
For other types of data ($\tau$ data especially) the
change can be larger. This uncertainty is a reflection of the mixture 
of contributions at NLO.

In the proposed model the analysis is unambiguous and straightforward.
I present different contributions separately for a detailed comparison with
the results from ref.~\cite{krause}.

For the vertex type contributions I use the
explicit analytical formulae in the leading order of
the mass expansion as they are given in ref.~\cite{samuel}. 
The exact expressions are presented in ref.~\cite{barbieri}.
The analytical expression for the contribution 
of a fermion with mass $m_q$ without any group factors
reads
\be
a_\mu^{\rm ferm}({\rm ver}; m_q)
=-\frac{2}{3}\left(\frac{m_\mu}{m_q}\right)^2\left(
-\frac{2689}{5400}+\frac{\pi^2}{15}
+\frac{23}{90}\ln\frac{m_q}{m_\mu}\right)\left(\alp\right)^3
\ee
that leads to a numerical result for the light mode contribution
in the model
\be
\label{vermassexp}
a_\mu^{\rm mod}({\rm ver;NLO; uds})=-172\times 10^{-11}\, .
\ee
A more accurate evaluation (using numerical integration with the kernel
given up to the third order in the mass expansion from ref.~\cite{krause}) 
gives for the contribution of light modes
\be
a_\mu^{\rm mod}({\rm ver;NLO; uds})=-188\times 10^{-11}\, .
\ee
The difference with the result obtained by using only the first term of mass
expansion given in eq.~(\ref{vermassexp})
is of order 10\%. It is smaller
than one could expect from the numerical value
of the expansion parameter $(m_\mu/m_q)^2=(0.106/0.179)^2=0.36$.
For the $c$-quark contribution one finds
\be
a_\mu^{\rm mod}({\rm ver;NLO; c})=-4\times 10^{-11}
\ee
while the $b$-quark contribution is small
\be
a_\mu^{\rm mod}({\rm ver;NLO; b})=-0.2\times 10^{-11}\, .
\ee
The total vertex contribution computed in the model 
\be
a_\mu^{\rm mod}({\rm ver;NLO})=-192\times 10^{-11}
\ee
should be compared with the result of the data-based analysis 
from ref.~\cite{krause}
\be
a_\mu^{\rm ref.~\cite{krause}}({\rm ver;NLO})
=-211(5)\times 10^{-11}\, .
\ee

Next check of the model is done for a mixed contribution 
of the lepton-hadron type.
This contribution contains the electron
and $\tau$-lepton loops and depends on three masses
$m_\mu$, $m_q$, and $m_e$ or $m_\tau$.
For fermions without group factors the contribution 
is given by the integral representation
\be
a_\mu^{\rm ferm}({\rm db;f_1\& f_2})=\left(\alp\right)^3
\int_0^1 dx (1-x) \pi(x,m_{f_1}) \pi(x,m_{f_2})\, .
\ee
For the combined contribution of light modes with the electron loop
one has
\be
\label{elpol}
a_\mu^{\rm mod}({\rm db;NLO;e\& uds})=105\times 10^{-11}
\ee
and with the $\tau$-lepton loop
\be
\label{taupol}
a_\mu^{\rm mod}({\rm db;NLO;\tau\& uds})=0.05\times 10^{-11}\, .
\ee
The contribution of heavy modes is only visible 
for the combined insertion of the $c$-quark loop and the electron 
loop
\be
\label{elcquapol}
a_\mu^{\rm mod}({\rm db;NLO;e\& c})=1.1\times 10^{-11}\, .
\ee
The results given in eqs.~(\ref{elpol}-\ref{elcquapol})
are in good agreement with the data-based estimate 
\be
a_\mu^{\rm ref.~\cite{krause}}({\rm lept\&had;NLO})
=107(2)\times 10^{-11}\, .
\ee
Next comes the contribution from the 
reiteration of hadronic insertions.
The light modes give
\be
\label{dbuds}
a_\mu^{\rm mod}({\rm db;NLO;uds\& uds})=3\times 10^{-11}\, .
\ee
The combination of the $c$-quark insertion with the light mode loops gives 
\be
\label{dbcuds}
a_\mu^{\rm mod}({\rm db;NLO;c\& uds})=0.1\times 10^{-11}
\ee
while the contribution of the two $c$-quark insertions is negligible.  
The results of the model from eqs.~(\ref{dbuds},\ref{dbcuds})
are in agreement with the data-based estimates
\be
a_\mu^{\rm ref.~\cite{krause}}({\rm had\&had;NLO})
=2.7\times 10^{-11}\, .
\ee
Thus one sees a good agreement of the model results with 
calculations based on data.
However, in the model there is a contribution 
which is missing in the explicit
calculations based on data as it is related to the internal structure of
the hadronic block. In the data-based calculation this contribution
is hidden in data while in the model 
it can explicitly be resolved as a correction to the one-particle
irreducible hadronic block. At the leading order of the mass ratio 
the analytical expression for this contribution without group factors
reads
\be
a_\mu^{\rm ferm}({\rm 4;NLO;}m_q)
=\frac{41}{486}\left(\frac{m_\mu}{m_q}\right)^2\left(\alp\right)^3\, .
\ee
The result for the light modes of the model is 
\be 
a_\mu^{\rm mod}({\rm 4;NLO; uds})= 25\times 10^{-11}
\ee
while the $c$-quark contribution is small
\be 
a_\mu^{\rm mod}({\rm 4;NLO; c})= 0.3\times 10^{-11}\, .
\ee

The result for the total NLO hadronic contribution of the vacuum 
polarization type is 
\be
a_\mu^{\rm mod}({\rm vac;NLO})=-58\times 10^{-11}\, .
\ee

The difference with eq.~(\ref{vacpol2})
comes mainly from two sources: the vertex type contributions and the new
term related to the one-particle irreducible hadronic block.
Both contributions are of the $(m_\mu/m_q)^2$ order that
explains the magnitude of the difference. 
All remarks about the double counting in the data-based approach apply here.
Only first few terms of expansions in the mass ratio $(m_\mu/m_q)^2$ 
were used for numerical estimates that provided a sufficient 
accuracy. 

Thus, the model reproduces rather accurately 
the results for the NLO hadronic contributions
found in the data-based analysis 
for the graphs related to vacuum polarization. 
This has been expected as these 
results are obtained by the integration of the two-point
function with the NLO kernel. 

The next try for the model is the computation of the 
light-by-light contribution which is given by the four-point correlator.  
The analytical expression for a contribution of the 
fermion without group factors 
through the $(m_\mu/m_q)^4$ order reads~\cite{lblanal}
\[
a_\mu^{\rm ferm}({\rm lbl;NLO;}m_q)=\left(\alp\right)^3\left\{
\left(\frac{m_\mu}{m_q}\right)^2
\left(\frac{3}{2}\zeta(3)-\frac{19}{16}\right)
\right.
\]
\be
\left.
+
\left(\frac{m_\mu}{m_q}\right)^4\left(
-\frac{161}{810}\ln^2\left(\frac{m_q}{m_\mu}\right)
-\frac{16189}{48600}\ln\left(\frac{m_q}{m_\mu}\right)
+\frac{13}{18}\zeta(3)-\frac{161}{9720}\pi^2
-\frac{831931}{972000}\right)\right\}\, .
\ee
With this formula one finds the value for the light modes
\be
a_\mu^{\rm mod}({\rm lbl;NLO;uds})
=140.5\times 10^{-11}
\ee
and one for the $c$ quark
\be
a_\mu^{\rm mod}({\rm lbl;NLO;c})=2\times 10^{-11}\, .
\ee
The total light-by-light contribution predicted by the model
\be
\label{modlbl0}
a_\mu^{\rm mod}({\rm lbl;NLO})=140.5+2=143\times 10^{-11}
\ee
is different from the number used in the literature~\cite{czamar}
\be
\label{litlbl}
a_\mu^{\rm had}({\rm lbl;standard})=-85(25)\times 10^{-11}\, .
\ee
I postpone a discussion of this point till sect.~(\ref{sect:discussion}).

Thus, the NLO hadronic contribution obtained in the model reads
\be
\label{totalsym}
a_\mu^{\rm mod}({\rm NLO})=(-58+143)\times 10^{-11}=85\times 10^{-11}\, .
\ee
It agrees with the present experimental result from
eq.~(\ref{nlopres}) which we repeat here 
\[
a_\mu^{\rm had}=(191\pm 175)\times 10^{-11}\, .
\]
The agreement with the future experimental result for MAMM
depends on a possible change of the mean value of $a_\mu^{\rm exp}$
as one sees from eq.~(\ref{futureNLOexp}). 

The prediction of the NLO hadronic contribution obtained in the model
is fairly sensitive to the
numerical value of the mass parameter for the light modes.
This numerical value is however strictly determined by the LO result. 
To check how sensitive to the details of the model the results are 
I introduce a mass difference
between $s$ and $u,d$ quarks ($SU(3)$ flavor violation in the mass
sector in the approximation of exact isotopic invariance).
I write $m_s=m_q+0.18~{\rm GeV}$ 
with $0.18~{\rm GeV}$ being the value of the running mass for 
the strange quark. Then one finds
\be
m_q=166\pm 1~{\rm GeV}
\ee
with 
\be
\label{vacpoludsback1}
a_\mu^{\rm mod1}({\rm LO; uds})(m_q=166\pm 1~{\rm MeV})
=(6928\pm 71)\times 10^{-11}\, .
\ee
The prediction of the NLO contribution in this case as compared to the
$SU(3)$ symmetric one is 
\be
a_\mu^{\rm mod1}({\rm NLO})-a_\mu^{\rm mod}({\rm NLO})
=(2+4+14)\times 10^{-11}
\ee
where the first term comes from vertex corrections, the second term comes
from insertions into the photon propagator and
the last term comes from the light-by-light graphs.
The result is fairly stable. Finally, the model with 
$SU(3)$ flavor violation in the mass sector gives the NLO hadronic
contribution to MAMM 
\be
\label{nonsym3}
a_\mu^{\rm mod1}({\rm NLO})=105\times 10^{-11}
\ee
which is rather close to the prediction of the model 
with $SU(3)$ symmetric mass arrangement from eq.~(\ref{totalsym}).

One could consider an even more sophisticated model including a
violation of the isotopic invariance
by using different masses for $u$ and $d$ quarks. An additional
uncertainty emerges from the errors in the numerical value for the
$c$-quark mass. By using the $\overline {\rm MS}$ mass around
$1.3~{\rm GeV}$ for the $c$-quark one could enhance its LO 
contribution by about 50\% (a leading order rescaling factor is 
$(m_c({\rm pole})/m_c(\overline {\rm MS}))^2=(1.6/1.3)^2=1.5$). 
Within the proposed model the use of the pole mass of the
heavy quark looks more natural while an account of the difference
between the numerical values for the pole and $\overline {\rm MS}$ 
masses is beyond the accuracy of the approximation used for heavy
quarks. It can readily be done
since the contribution of heavy quarks is perturbative and corrections
in the strong coupling constant can reliably be found.

\section{Discussion}
\label{sect:discussion}
The underlying idea of the presented analysis is to introduce a framework 
for computing the NLO hadronic contributions to MAMM using the LO information.
Presently the results for the light-by-light contribution 
that is the most interesting term at NLO are available analytically 
for fermions that dictates the choice of the model from the technical
point of view almost uniquely. Thus, a model of massive quarks 
with the EM interaction emerges as a suitable candidate. 
It is not an approximation for QCD as a gauge model with constituent
quarks. It is just a bridge from LO to NLO results for hadronic
contributions to a particular observable. 
Note that for another important parameter of the standard model -- the
running EM coupling constant at the scale of the Z-boson mass --
there is no possibility to use such kind of a model as there is
no important next-to-leading order terms to compute.
Calculations for the infrared (IR) sensitive observables 
using constituent quarks with masses around 
300--500 MeV as the only IR scales are unjustified in pQCD in general 
since the higher order corrections in the strong coupling constant 
cannot be found. Also the introduction of finite masses for the light 
quarks explicitly violates chiral invariance
which is a well established symmetry of the light hadronic sector.
In this sense the approximation for QCD with constituent quarks cannot
be considered as a reasonable general framework. In the high energy limit 
the massless approximation for strong interactions is perturbative
and quite precise. This means that high energy contributions to MAMM 
can be represented by almost any model that satisfies the duality
constraints. In this sense the fermionic 
model fits the standard approximation for large energies.
However, the main contribution to MAMM comes from the IR region
where there is no sensible approximation for strong interactions
deduced from QCD. Therefore the necessary characteristics 
of the strong interaction amplitudes relevant for the computation
of MAMM has to be extracted from data.
The first amplitude that emerges is the two-point correlator 
that is given by a single function of one variable with simple 
analytic properties (see eqs.~(\ref{pi2},\ref{vacpolin})). 
For computing MAMM one need not know the point-wise behavior of 
the spectrum but
only the integral over all energies with some enhancement of the
threshold region. A model of massive fermions is then well suitable to
fit this integral over data. When hadrons are introduced into the threshold 
IR region to fit experiment the effective masses of quarks
increase. Therefore an account of low-energy
hadronization for the two-point function entering MAMM
is achieved by introducing an explicit cut in energy in the sum over
the states. In practice, at the leading order 
the hadron contributions are represented by the pion
with an EM interaction of the form $e j^\pi_\mu  A^\mu$
where at the leading order $j^\pi_\mu 
=i(\pi^+\stackrel{\leftrightarrow}{\partial}_\mu \pi^-)$.
The inclusion of pions leads to the
scalar type of the spectrum near the threshold
\be
\rho_\pi(s)=\frac{1}{12}\sqrt{1-\frac{4 m_\pi^2}{s}}
\left(1-\frac{4 m_\pi^2}{s}\right)
\ee
instead of the fermionic form given in eq.~(\ref{fermspect}).
Furthermore, the fermionic contributions can be moved to higher energies
by using vector mesons. In the vector meson dominance 
model one identifies the EM current with the canonically normalized
elementary $\rho$-meson field $\rho_\mu$ through the relation
$j_\mu^{had}=f_\rho \rho_\mu$. Here $f_\rho$ gives a form factor related to 
the leptonic width of the $\rho$-meson. 
Because of the nature of the MAMM observable this contribution 
can well be represented by fermions already as it resides at a rather
large scale. This hadronization picture is transparent for 
the two-point function which is sufficient for the LO analysis. 
At NLO an hadronization procedure for the four-point function is necessary.
Within a hadron picture of the low-energy spectrum the most
important contribution to MAMM comes from pions.
To quantitatively handle contributions from the four-point function 
a quantum field model for pions given by the Lagrangian
\be
L_{low energy}=|D_\mu\pi|^2-m_\pi^2 \pi^2, 
\quad D_\mu=\partial_\mu-i e A_\mu
\ee
is introduced.
This model generates vertices that allow to compute
the pion contribution to the four-point hadronic EM current 
correlator that enters the light-by-light diagram 
explicitly. The high energy contribution of this model 
should then be replaced by the standard quark contributions. 
In the pure fermionic model with the small effective mass this
replacement is effectively made at rather low energies that makes 
the separate contribution of pions small or even vanishing.
Thus, the hadronization procedure of the model is realized through
light massive quarks rather than real hadrons. Note that the
hadronization picture need not be universal for all strong interaction
processes but can specially be tailored for a given observable.

The results for MAMM related to the two-point function which
have been obtained in data-based analysis are well reproduced by 
the model with the mass of the light fermion around the pion mass. 
Using the model prediction for the light-by-light graph 
I find an agreement of the NLO hadron contribution to MAMM
with experiment.
The results for the light-by-light graph
in the pion model are available numerically. 
In the absence of analytical expressions for the light-by-light
contributions in the pion model I could not quantitatively check 
how fermionic contributions replace the pion ones when the effective 
fermion mass decreases. However, it seems probable that 
the explicit inclusion of the pion contributions in the framework of 
the present model will shift the effective mass of light quarks larger.

The fermionic model gives a smooth spectrum at low energies.
An important question is whether such a smooth spectrum is a
reasonable approximation for computing MAMM. 
In the two-point correlator there are no
resonances in the relevant region. The contribution
of the $\rho$-meson is located at relatively large scales.
In the axial channel, for instance, the situation is different because of the
presence of the pion resonance and probably
the model with massive quarks would not fit. 
Note, however, that as soon as the pion is considered to be 
massive (not a pure Goldstone mode)
the chiral invariance is explicitly broken that makes quarks 
massive as well (or vice versa).
Note also that a model can be suited for a description of a
specific observable and need not give a universal approximation
of any Green function.
For instance, in the axial-vector two-point 
correlator the projection related to spin-one
particles contains only massive resonances and the spectrum can 
well be approximated by a fermionic model without the pion pole.
For the four-point functions the situation is more complicated though.
In the literature there are models where the four-point function at
low energies is represented through 
the elementary fields of neutral pseudoscalar bosons
in order to compute contributions of the light-by-light graphs
to MAMM. 
The representation employs the neutral pion
contribution to the four-point function through the iteration of 
an effective Lagrangian for the interaction of the neutral pion
$\pi^0$ with photons due to the Abelian anomaly in axial current
(as a review see ref.~\cite{pseudo}). 
The result for the light-by-light contribution obtained using the
neutral pion dominance eq.~(\ref{litlbl})
is different from one obtained in the present model eq.~(\ref{modlbl0}).
In general, the reduction of the four-point amplitude of the hadronic
EM currents to a two-point correlator of axial currents uses
the operator product expansion at small distances
\be
iTj_\mu^{had}(x)j_\nu^{had}(-x)|_{x\to 0}
= \varepsilon_{\mu\nu\omega\lambda}x^\omega j_5^\lambda(0) C(x^2)+\dots
\ee
where $C(x^2)$ is a coefficient function of the local operator 
$j_5^\lambda(0)$ which has quantum numbers of axial current 
(e.g.~\cite{ope}). 
In other words the combination of two hadronic EM currents of the form 
\be
\varepsilon_{\mu\nu\omega\lambda}\xi^\lambda
T j_\mu^{had}(x+\xi)j_\nu^{had}(x-\xi) F(\xi^2)
\ee
taken at small $\xi$ with some form factor $F(\xi^2)$
may act in some applications 
as a local axial current that can serve as an interpolation field for 
the pion. Thus, this combination can be replaced by a fundamental 
pion field in a hadronization procedure. 
This kind of factorization for the four-point amplitude 
is valid for $\gamma \gamma \to \gamma \gamma$ scattering in a specific
region of the phase space in kinematic variables where all 
three external momenta are essential.
In other regions of the phase space the saturation of the 
scattering amplitude with the pion pole contribution is invalid.
The projection of the four-point function 
that emerges in the light-by-light graphs for the MAMM calculation
has the form
\be
\label{lblproj}
\int dx x^\delta
\langle Tj_\mu^{had}(x)j_\nu^{had}(y)
j_\al^{had}(z)j_\beta^{had}(0)\rangle\, .
\ee
In momentum space this projection depends on two external momenta only
as the third momentum is set to zero after differentiation according 
to the definition of MAMM. In the neutral pseudoscalar model the
projection of the four-point function given in eq.~(\ref{lblproj})
is saturated by the contribution of the neutral pion
that seems to be invalid in the kinematical region
relevant for MAMM computation. 
In the absence of the neutral pion pole contribution 
in the hadronization picture for the light-by-light graph
the fermionic model can be used for its computation
on the same footing as it was used for vacuum polarization graphs. 
In fact, the neutral pion 
contribution gives the major difference with the present analysis 
based on the fermionic model. However, the corresponding contribution 
of the neutral pion to the projection of the four-point function 
emerging in the photon propagator is usually not considered. 
In other words, the neutral pion approximation for the four-point function
should also be taken into account in eq.~(\ref{proj2phot})
as it is accounted for in eq.~(\ref{lblproj}).
If the neutral pion contribution to eq.~(\ref{proj2phot})
does not vanish by some symmetry considerations
it can lead to a cut starting from the pion mass square $m_\pi^2$
that seems to contradict the threshold behavior of the spectrum
known in $e^+e^-$ annihilation. This calls for a necessity
to evaluate the validity of the neutral pion dominance 
model for the four-point function in the kinematical region relevant 
for computing the NLO hadronic contribution to MAMM. 

Despite the fact that the fermionic model with the mass $m_q=179~{\rm MeV}$
predicts the value for the NLO hadronic contribution to MAMM in
agreement with experiment there remains a disturbing feeling that 
this prediction is obtained within an unrealistic approximation for
strong interactions and, therefore, cannot be taken seriously.
A historic reminiscence may be appropriate here. A century ago thinking about
the light as existing in the form of discrete portions -- photon quanta --
was rather disturbing for classical physics. However, the
quantum representation allowed for the quantitative 
explanation of experimental facts on photoeffect and 
black body radiation. It did not change the description of electromagnetic 
phenomena insensitive to the quantum nature of the light.
It may happen that the muon anomalous magnetic moment is
sensitive to the contribution of all hadrons in a way it would be sensitive
to that of free fermions with an appropriate mass
which is a standard realization of duality concept.
The direct application of this concept to a particular case  
of MAMM looks suspicious because the IR region is explicitly 
involved in the analysis and the results depend strongly on 
the numerical value of the effective quark mass which happens 
to be rather small. The model, however, is only designed for computing 
the NLO hadronic contribution 
to MAMM using the LO result as input.
This does not mean that this model approximation suited for computing MAMM
is in any sense a universal limit of QCD automatically
applicable to other observables.

\section{Conclusion}
A model for describing the NLO hadronic
contributions to the muon anomalous magnetic moment
is proposed. The model contains a single parameter 
that is fixed from the experimental result for the LO hadronic
contribution to MAMM. The model describes the NLO hadronic
contributions of the vacuum polarization type in agreement with
existing estimates. 
However, it predicts a numerical value for the light-by-light 
contribution which is different from 
one used in the literature that considerably changes the 
prediction of the total NLO hadronic contribution to MAMM.
The prediction of the model agrees with 
the present experimental value for MAMM.
A resolution of the contradiction between 
the estimates for the NLO hadronic contribution, 
or rather for the light-by-light contribution, obtained in the present
model and existing in the literature 
could help in verifying the validity of the standard model.

\vspace{5mm}
\noindent
{\large \bf Acknowledgments}\\[2mm]
I thank A.I.Davydychev and J.G.K\"orner for interesting discussions.
This work is partially supported by Russian Fund for Basic Research 
under contract 99-01-00091 and 01-02-16171. 
My present stay at Mainz University is made possible by a grant from DFG.


\begin{thebibliography}{99}
\bibitem{czamar}A.Czarnecki and W.J.Marciano, 
Phys. Rev. {\bf D64}, 013014 (2001) 
%%CITATION = PHRVA,D64,013014,2001;%%
\bibitem{tenthQED}T.Kinoshita, B.Nizic, Y.Okamoto,
Phys. Rev. {\bf D41}, 593 (1990)
%%CITATION = PHRVA,D41,593,1990;%%
\bibitem{qedcont}P.J.Mohr and B.N.Taylor, 
Rev. Mod. Phys. {\bf 72}, 351 (2000)
\bibitem{sakurai}C.F.Cho, J.J.Sakurai, 
Phys. Lett. {\bf  B30}, 119 (1969); \\
%%CITATION = PHLTA,B30,119,1969;%%
J.J.Sakurai, Phys. Lett. {\bf B46}, 207 (1973)
%%CITATION = PHLTA,B46,207,1973;%%
\bibitem{leadorder}S.Brodsky, E. de Rafael, 
Phys. Rev. {\bf 168}, 1620 (1968)
%%CITATION = PHRVA,168,1620,1968;%%
\bibitem{kinohad}T.Kinoshita, B.Nizic, Y.Okamoto,
Phys. Rev. {\bf D31}, 2108 (1985)
%%CITATION = PHRVA,D31,2108,1985;%%
\bibitem{davhock}R.Alemany, M.Davier, A.H\"ocker, 
Eur. Phys. J. {\bf C2}, 123 (1998);\\
%%CITATION = EPHJA C2,123,1998;%%
M.Davier, A. H\"ocker,  Phys. Lett. {\bf B435}, 427 (1998)
%%CITATION = PHLTA,B435,427,1998;%%
\bibitem{jeger}S. Eidelman and F.Jegerlehner,
Z. Phys. {\bf C67}, 585 (1995); F.Jegerlehner, in ref.~[1]
%%CITATION = ZEPYA,C67,585,1995;%%
\bibitem{eucl}J.G.K\"orner, A.A.Pivovarov and K.Schilcher,
Eur. Phys. J. {\bf C9}, 551 (1999);\\
%%CITATION = EPHJA C9,551,1999;%%
A.A.Pivovarov,
``Running electromagnetic coupling constant: low-energy normalization
and the value at M(Z).''
MZ-TH-00-51, Nov 2000. 53pp. 
[hep-ph/0011135]
%%CITATION = HEP-PH 0011135;%%
\bibitem{PDG}C.Caso {\it et al.,} (Particle Data Group),
Eur. Phys. J. {\bf C3}, 1 (1998)
%%CITATION = EPHJA C3,1,1998;%%
\bibitem{mb}J.H.K\"uhn, A.A.Penin and A.A.Pivovarov,
Nucl. Phys. {\bf B534}, 356 (1998);
%%CITATION = NUPHA,B534,356,1998;%%
A.A.Penin and A.A.Pivovarov,
Phys. Lett. {\bf B435}, 413 (1998);
%%CITATION = PHLTA,B435,413,1998;%%
Nucl. Phys. {\bf B549}, 217 (1999)
%%CITATION = NUPHA,B549,217,1999;%%
\bibitem{krause}B.Krause, Phys. Lett. {\bf B390}, 392 (1997)
%%CITATION = PHLTA,B390,392,1997;%%
\bibitem{samuel}M.A.Samuel and G.Li, Phys. Rev. {\bf D44}, 3935 (1991)
%%CITATION = PHRVA,D44,3935,1991;%%
\bibitem{barbieri}R.Barbieri and E.Remiddi, 
Nucl. Phys. {\bf B90}, 233 (1975)
%%CITATION = NUPHA,B90,233,1975;%%
\bibitem{lblanal}S.Laporta and E.Remiddi, 
Phys. Lett. {\bf B301}, 440 (1993)
%%CITATION = PHLTA,B301,440,1993;%%
\bibitem{pseudo}M.Hayakawa and T.Kinoshita, 
Phys. Rev. {\bf D57}, 465 (1998)
%%CITATION = PHRVA,D57,465,1998;%%
\bibitem{ope}A.A.Pivovarov, 
Phys. Rev. {\bf D47}, 5183 (1993);
%%CITATION = PHRVA,D47,5183,1993;%%
A.A.Penin and A.A.Pivovarov,\\
Nucl. Phys. {\bf B550}, 375 (1999);
%%CITATION = NUPHA,B550,375,1999;%% 
Phys. At. Nucl. {\bf 64}, 275 (2001)
%%CITATION = HEP-PH 9904278;%%

\end{thebibliography}
\end{document}